\documentclass[graybox]{svmult}

\usepackage{type1cm}        
\usepackage{makeidx}         
\usepackage{graphicx}        
\usepackage{multicol}        
\usepackage[bottom]{footmisc}

\usepackage{}

\usepackage{newtxtext}       %
\usepackage[varvw]{newtxmath}       

\makeindex             

\begin{document}

\title*{Fluctuations of Local Patterns in 
Non-Periodic Uniquely Ergodic Systems}
\author{Jacek Mi\c{e}kisz \orcidID{0000-0001-6651-8951}} 
\institute{Jacek Mi\c{e}kisz \at University of Warsaw, Institute of Applied Mathematics and Mechanics, Banacha 2, 02-097 Warsaw, Poland \email{miekisz@mimuw.edu.pl}}

%
%
\maketitle

With all the best wishes to Vitaly. I am very happy that our trajectories crossed much earlier than the ergodic theory would predict. 
\vspace{5mm}

\abstract{We study various families of non-periodic configurations of symbols assigned to vertices of regular graphs. We restrict ourselves to such systems where considered configurations support a uniquely ergodic measure. Symbols can be interpreted as particles and we are concerned with configurations which minimize the interaction energy of particles, the so-called ground-state configurations. One of the main questions then is 
whether ground-state configurations are stable with respect to small perturbations of interactions. 
Such stability may be interpreted as a sort of rigidity of non-periodic structures such as tilings, substitution systems, 
Sturmian systems, and other uniquely ergodic non-periodic systems. We introduce a property of bounded fluctuations of finite patterns of non-periodic configurations, the so-called Strict Boundary Property. We discuss how relevant this property is for stability of non-periodic structures. Examples of Robinson's tilings, Thue-Morse, and Sturmian systems are presented. Some open problems are formulated.}

\section{Introduction}
\label{sec:1}

Motivation of our research comes from quasicrystals, state of matter which is ordered but does not possess a structure of periodic crystals. They were discovered in 1982 by Dan Shechtman \cite{shechtman}. However, the study of non-periodic systems is much older. It might be traced back to David Hilbert who presented his famous 23 problems in 1900. The second part of the 18-th problem can be phrased in the following way: does there exist a polyhedron which can cover the space but only in a non-periodic way? Then comes Hao Wang, in 1961 he translated the Hilbert problem for domino players \cite{wang} and conjectured that if we can tile the plane with square dominoes such that colors of neighboring sides are the same, we can also tile it in a periodic way. The first counter-example, consisting of over 20 thousand square-like tiles, was constructed in 1966 by Richard Berger \cite{berger}. Then in 1971 the number of tiles was greatly reduced to 56 by Raphael Robinson \cite{robinson}. In 1974, Roger Penrose constructed two polygons, famous kite and dart (they are not square-like Wang tiles), which cover the plane only in a non-periodic way \cite{penrose}. 

Very recently there were presented two constructions of "einstein" - an affirmative answer to the Hilbert question. First, a 13-sided polygon called the Hat was constructed by David Smith, Joseph Samuel Myers, Craig S. Kaplan, and Chaim Goodman-Strauss \cite{hat}. However, to tile the plane one has to use also its mirror image. Then the same authors gave us in \cite{spectre} the Spectre, only its rotations are needed to tile the plane. Hence, finally the Hilbert problem was solved.

The history of one-dimensional non-periodic sequences is very old. Thue-Morse sequences were introduced by Axel Thue in 1906 in the study of combinatorics on words \cite{Thue} and rediscovered by Marston Morse in 1921 in the context of differential geometry and geodesics \cite{Morse}. They were studied by Michael Keane in symbolic dynamical systems \cite{keane}. We should also mention here Sturmian systems resulting from rotations on a unit circle by irrational numbers \cite{sturmian1,sturmian2}. Connection between one-dimensional substitution systems like Thue-Morse and Fibonacci (an example of a Sturmian system based on rotations by the golden ratio) was put forward by Shahar Mozes \cite{mozes} in constructions of tilings of the plane by Wang square-like tiles realizing substitution rules. 

From our ergodic point of view, all these examples have one thing in common. In each of them, all non-periodic structures, non-periodic configurations on ${\mathbb Z}^{d}$, support a uniquely ergodic measure with respect to translations. More precisely, the closure in the product topology of an orbit by translations of any of them is a minimal, uniquely ergodic subshift with minimality giving equal patterns, unique ergodicity equal frequencies, so that the orbit closure coincides with the support of the unique invariant measure.

Our point of view is to see uniquely ergodic systems as results of minimization of certain functionals on the set of measures. Such an idea functions in the literature as an ergodic optimization \cite{opterg,typical}. Certain results concerning generic properties of optimizing measures were obtained. It was shown that in the space of infinite-range but summable interactions (the energy of every particle is finite), non-periodic ergodic measures which are non-mixing so possess some sort of order, form a generic set in the space of interactions - a set of the second Baire category \cite{jaradin2,jaradin3}.

Our approach here is a constructive one. For a given uniquely ergodic measure we try to construct simple energy functionals, hamiltonians which describe interactions between particles corresponding to tiles in two-dimensional tilings or symbols in one-dimensional dynamical systems. Configurations which minimize the energy of interacting particles are called ground-state configurations, the ergodic probability measure which supports them is called a ground-state measure.  

In tiling systems, we assign to particles a positive energy, say $1$, if corresponding tiles do not match, otherwise the energy is $0$. In general systems of finite type, we assign positive energies to finite number of finite forbidden patterns (which define such a system). In this way we construct classical lattice-gas models (see a recent monograph on lattice models of interacting particles \cite{velenik} and a recent survey on lattice models of quasicrystals \cite{quasireview}). First examples were based on Robinson's tilings \cite{radinquasi,manyphases,sequenceoftemp}. They constituted counterexamples to the old Crystal Problem - the long-standing but never rigorously established expectation among physicists that at low enough temperatures and  high enough pressures, any macroscopic matter should crystalize and therefore positions of atoms or molecules should be invariant by some non-trivial translations \cite{simon}. Classical-lattice gas models based on Robinson's tilings do not have periodic ground states which minimize the energy of the system at zero temperature and hence they show that expectation of physicists fails in general. 

It is known that any one-dimensional classical lattice-gas model with finite-range interactions have at least one periodic ground-state configuration \cite{ban,radinschulman,thirdlaw}. Therefore to force non-periodicity we have to use infinite-range interactions with the proviso that the interaction
energy decays fast enough with distance for the total energy of each particle to remain finite (summability).

Our goal here is to construct simple interactions for which non-periodic ground-state configurations form a support of a uniquely ergodic measure (there are no periodic ground-state configurations). To mimic constructions based on two-dimensional systems of finite type, we are looking for the minimal set of forbidden patterns (necessarily an unbounded one) which characterize non-periodic configurations in the support of a given ergodic measure. Such systems might be called systems of minimal infinite type. We will discuss constructions related to Thue-Morse systems and Sturmian ones in Section 3. 

The fundamental question is whether non-periodic uniquely ergodic ground-state measures are stable with respect to small perturbations of interactions between particles. We argued in \cite{sbp1,sbp2} that bounded fluctuations of finite patterns, called by us the Strict Boundary Property, is essential to such stability.
 
\section{Strict Boundary Property in two-dimensional systems}
\label{sec:2}

Here we discuss two-dimensional dynamical systems of finite type, in particular tiling systems.

Let $\Omega = \{1,...,n\}^{{\mathbb Z}^{2}}$ be a space of configurations equipped with the product topology. We will be concerned with non-periodic uniquely ergodic systems $({\mathbb X}, \mu)$ of finite type, called subshifts of finite type (SFT), where $\mu$ is the unique probability ergodic measure supported by ${\mathbb X} \subset \Omega$. It means that ${\mathbb X}$ contains only non-periodic configurations (not invariant under any lattice translations) and it is defined by the absence of a finite number of certain local configurations, called local patterns, $p  \in \{1,...,n\}^V$ for some finite $V \subset {\mathbb Z}^{2}$. By $\omega_{p}$ we denote the frequency of $p$ in $\mu$. Let $\Lambda \subset {\mathbb Z}^{2}$  be a finite set. Here $X(\Lambda) \in \{1,..., n\}^{\Lambda}$ denotes a locally admissible configuration on $\Lambda$, i.e. one containing
no forbidden patterns. Denote by $n_{p}(X(\Lambda))$ the number of translates of $p$ in $X(\Lambda)$. Let us notice that in tiling systems, forbidden patterns are pairs of nearest-neighbor tiles that do not match.
\vspace{2mm}

\begin{definition} A uniquely ergodic system $({\mathbb X}, \mu)$ 
has {\bf Strict Boundary Property} for a pattern $p$ if there exists 
$C_{p}  < \infty$ such that

$$|n_{p}(X(\Lambda)) - \omega_{p}|\Lambda|| < C_{p} \delta \Lambda$$

for every $\Lambda$ and every admissible $X(\Lambda)$, where $|\Lambda|$ is the number of lattice points in $\Lambda$ and $\delta \Lambda$ is the length of its boundary. The system has the Strict Boundary Property if it has the property for every finite pattern \cite{sbp1,sbp2}.
\end{definition}

\noindent {\bf Open Problem 1} 
\vspace{3mm}

\noindent Does there exists a uniquely ergodic non-periodic system of finite type with Strict Boundary Property?
\vspace{3mm}

\noindent Such property is also called the rapid convergence to equilibrium of frequency of patterns \cite{peyriere,gambaudo}. This is the requirement 
that the number of any finite pattern in a finite lattice subset can fluctuate at most proportional to its boundary. One can easily generalize this definition to tilings with not necessarily square-like tiles, like Penrose tilings. Some other related concepts were introduced in different contexts, see \cite{aizenman} and papers on hyperuniformity \cite{torquato1,torquato2,maher}. 

Now we would like to present systems of finite type as classical lattice-gas models with finite-range interactions between particles. Symbols of dynamical systems, that is, elements if its alphabet, are identified with particles which are placed at lattice sites. We assign a positive energy to forbidden patterns of particles and zero otherwise. In this way we introduced interactions between particles. Formally we define a hamiltonian
\begin{equation}
H(X) = \sum_{p} 1,
\end{equation}
where $X \in \Omega$ and the summation is with respect to all forbidden patterns in $X$. 

In tiling examples, these are nearest and next-nearest neighbor interactions on ${\mathbb Z}^2$. Then configurations of particles without forbidden patterns have zero energy, they are called ground-state configurations of the above described hamiltonian. The uniquely ergodic measure which supports them is called a ground-state measure. The following theorem connects the Strict Boundary Property with the stability of ground-state configurations \cite{sbp1}.
\begin{theorem}
A unique ground-state measure of a finite-range interaction is stable against small perturbations of interactions of range smaller than $r$ if and only if it has the Strict Boundary Property for patterns of diameters smaller than $r$, diameter of a pattern is the maximal distance between its two sites.
\end{theorem}

\noindent It can be shown that a lattice-gas model based on Robinson’s tilings does not have this property \cite{unstable}.
There are arbitrary long sequences of tiles, called arrows, pointing in one direction such that if one changes them to tiles 
with arrows pointing in the opposite direction, one has mismatches (pairs of particles with the positive interaction energy) only at the end of sequences. One may therefore introduce a one-particle interaction (called by physicists a chemical potential) favoring tiles with an arrow in the opposite direction, that is assigning a negative energy to them. We have the following theorem \cite{unstable}.
\vspace{2mm}

\begin{theorem} The Robinson’s ground-state measure is not stable against an arbitrarily small one-particle interaction assigning a negative energy to one type of particles.
\end{theorem}

\noindent Instability of the Robinson's system was also discussed in \cite{gonsquas}. Positive answer to the Problem 1 would yield us a stable non-periodic ground-state measure - a microscopic model of a quasicrystal.

\section{One-dimensional uniquely ergodic systems}
\label{sec:3}

We have already pointed out that every one-dimensional lattice-gas model with finite-range interactions has at least one periodic ground-state configuration \cite{ban,radinschulman,thirdlaw}. Hence there are no one-dimensional non-periodic uniquely ergodic systems of finite type. We have to construct systems which are not SFT, correspondingly systems with infinite-range interactions. Below we present two examples of uniquely ergodic systems
which involve configurations of two symbols $+,-$ in the spin language or $0, 1$ in the particle language: Thue-Morse and Sturmian uniquely ergodic systems. We will study stability of their ground-state configurations 
with respect to small perturbations of interactions. However, we should be subtle here. One should understand "small" as a small norm in some space of interactions. The most natural choice is $l_{1}$ space of summable interactions which guarantees that the interaction energy of any particle with all others is finite. However, in such a space a tail of interactions can be arbitrary small and so we can put it to zero and in this way we are left with a finite-range interaction and we know that it has at least one periodic ground-state configuration. Therefore we conclude that one-dimensional non-periodic ground-state measures cannot be stable in this manner. Instead we will investigate whether they are stable 
against small finite-range perturbations. In the first place we study stability against one-site perturbations - a small chemical potential favoring the presence of particles $0$ (assigning a negative energy to $0$) or equivalently a small magnetic field favoring $+$ orientation of spins ($+$ has a negative energy $-h$, where $h>0$ is a magnetic field).
 
\subsection{Thue-Morse systems}
\label{subsec:1}

Thue-Morse sequences can be constructed in the following way. We place $+$ at the origin and successively perform the substitution $S$: $ + \rightarrow +-, - \rightarrow -+$.
In this way, we get a one-sided sequence $+--+ -++- -++- +--+ ...$, $\{X_ {TM }(i)\}, i \geq 0$. 
We define $X_{TM} \in \Omega = \{-, +\}^{Z}$ by setting 
$X_{TM}(i) = X_{TM} (-i-1 )$ for $i<0$.  
Let $G_{TM}$ be the closure of the orbit of $X_{TM}$ by the translation operator $T$, 
i.e., $G_{TM} = \{T^{n} (X_{TM}), n \geq 0\}^{cl}.$ It has been proved by Michael Keane that $G_{TM}$ 
supports exactly one translation-invariant probability measure $\mu_{TM}$ on $\Omega$ \cite{keane}.

It was shown by Gottschalk and Hedlund \cite{ghmorse} that Thue-Morse sequences are exactly those which do not contain any subword $BBb$ where $B$ is any word and $b$ is its first symbol. It was shown in \cite{tmhamiltonian} that the above characterization is equivalent to the condition which involves forbidden patterns at some infinite set of 4 lattice sites only. Namely, we cannot have two pairs at a distance $(2p+1)2^{r}$ such that a pair consists of the same two symbols at a distance $2^{r}$ for any $r, p \geq 0$. We may call such a system a uniquely ergodic system of a minimal infinite type. Now we can penalize forbidden patterns by assigning to them a positive energy which decays appropriately to zero with the growing pattern diameter. 

In \cite{tmhamiltonian} it was shown that $\mu_{TM}$ is the only ground-state measure of the following exponentially decaying four-spin interactions,

\begin{equation}
H_{TM} = \sum_{r=0}^{\infty} \sum_{p=0}^{\infty} H_{r,p},
\end{equation}

where

\begin{equation}
H_{r,p} = \sum_{i \in Z} J(r,p) (\sigma_{i} + \sigma_{i+2^{r}})^{2} (\sigma_{i+(2p+1)2^{r}} + \sigma_{i+(2p+2)2^{r}})^{2}
\end{equation}

and $\sigma_{i}(X)= X(i) \in \{-1, +1\}$, we identify $-1$ with $-$ and $+1$ with $+$.

It was shown recently that the Thue-Morse system is unstable against arbitrarily small perturbations of interactions \cite{jspnonstab}.

\begin{theorem}
The Thue-Morse ground state $\mu_{TM}$ is unstable against an arbitrarily small interaction that assigns negative energy to molecules consisting of two up or two down neighboring spins.
\end{theorem}

\subsection{Sturmian systems}
\label{subsec:2}

As in Section 3.1, we again consider sequences over a two-letter alphabet, here in the particle language $\{0, 1\}$, with $0$ and $1$ denoting an empty and an occupied site, respectively.

We will identify the circle ${\mathbb T}$ with ${\mathbb R}/{\mathbb Z}$ and consider an irrational rotation by $\varphi$ (which is given by translation on ${\mathbb R}/{\mathbb Z}$ by $\varphi \mod 1$).   

\begin{definition}
Given an irrational $\varphi\in {\mathbb T}$ and $x \in {\mathbb T}$ we say that $X\in\{0,1\}^{\mathbb Z}$ is generated by $(x,\varphi)$ if it is of the following form: 
\begin{equation*}
    X(n) = \begin{cases}
               0               & \text{when} \ x+n\varphi \in P \\
               1                & \text{otherwise}
           \end{cases}
\end{equation*}
where $P=[0,\varphi)$.
\end{definition}

We call such $X$ a Sturmian sequence corresponding to $(x,\varphi)$. 
Let $G_{St}$ be the closure of the orbit of $X$ by translations. 
It can be shown that $G_{St}$ supports exactly one translation-invariant probability measure $\mu_{St}$ on $\Omega$. 

Sturmian sequences can be characterized by the absence of certain finite patterns.

\begin{theorem}\label{forbidden}\emph{\cite[Theorem 4.1]{sturmiangs}}
Let $\varphi\in (\frac{1}{2},1)$ be irrational. Then there exist a natural number $m$ and a set $F\subseteq N$ of forbidden distances 
such that Sturmian sequences generated by $\varphi$ are uniquely determined by the absence of the following patterns: 
$m$ consecutive 0's and two 1's separated by a distance from $F$.
\end{theorem}

In the case of the Fibonacci system, where $\varphi = 2/(1 + \sqrt{5})$, $m=3$, and allowed distances are $d_{j}$ and $d_{j}+1$, $j=1,...$ where $d_{j}$ is the largest integer smaller than $j(2 +\varphi)$. Alternatively, Fibonacci configurations can be constructed by the substitution $0 \rightarrow 01, 1 \rightarrow 0$.

\begin{theorem}\label{sturmian_unstable}
Assume that $\varphi \in (0,1)$ is badly approximable and the interaction energy decays at least as fast as $1/r^{\alpha}$ with $\alpha>3$. 
Then Sturmian ground-state configurations generated by $\varphi$ are unstable with respect to an arbitrarily small chemical potential favoring the presence of particles, that is one-particle interaction assigning a negative energy to particles.
\end{theorem}

\noindent {\bf Open Problem 2}
\vspace{3mm}

\noindent Does there exist $\alpha > 1$ and $e < 0$ such that the Sturmian ground state with the interaction energy decaying as $1/r^{\alpha}$ is stable against a chemical potential - one-particle interaction assigning energy $e$ to particles?
\vspace{3mm}

\noindent Of course we would like to extend such a stability to all finite-range perturbations of interactions. 

\section{Discussion}
\label{sec:4}

We have discussed uniquely ergodic measures which support non-periodic configurations of symbols on ${\mathbb Z}^{d}, d=1,2$, which minimize simple hamiltonians - interactions between particles corresponding to symbols. They are called ground-state configurations. One of the main questions then is 
whether ground-state configurations are stable against small perturbations of interactions. Such stability may be interpreted as a sort of rigidity of non-periodic structures. 
We introduce a property of bounded fluctuations of finite patterns of non-periodic configurations, the so-called Strict Boundary Property. We discuss how relevant this property is for stability of non-periodic structures in uniquely ergodic systems. Examples of Robinson's tilings, Thue-Morse, and Sturmian systems are presented. Some open problems are formulated. We refer to \cite{sadun} for an extensive list of open problems, see also a recent monograph by Michael Baake and Uwe Grimm \cite{baake}.

The main question, from our point of view, can be formulated as follows: Does there exist a classical lattice-gas model with finite-range interactions and with a uniquely ergodic non-periodic ground-state measure which is stable with respect to finite-range perturbations of interactions and with a non-periodic Gibbs state at low temperatures? 

The first part of the above question was formulated in Section 2 as Open Problem 1. Open Problem 2 in Section 3 concerns one-dimensional non-SFT models hence lattice-gas models with infinite-range summable interactions and it asks about stability only against finite-range interactions. Both problems concern systems of interacting particles at the zero temperature. The second part of the main question means stability of the ground-state measure against thermal motions of particles at low positive temperatures. Then one has to minimize the free energy of the system instead of just the energy. Measures on the space of configurations which minimize the free energy are called Gibbs states, they describe equilibrium behavior of systems of interacting particles at non-zero temperatures. For details we refer to a recent survey \cite{quasireview}. Construction of non-periodic Gibbs measure based on non-periodic uniquely ergodic ground-state measure would give us a first microscopic toy model of a thermodynamically stable quasicrystal. 

\begin{acknowledgement}
I would like to thank the National Science Centre (Poland) for a financial support under Grant No. 2016/22/M/ST1/00536, Charles Radin and Aernout van Enter for many years of collaborations.
\end{acknowledgement}


\begin{thebibliography}{99}

\bibitem{shechtman} D. Shechtman, I. Blech, D. Gratias, and J.W. Cahn,  
Phys. Rev. Lett. 53: 1951-1953 (1984).

\bibitem{wang} H. Wang, Proving theorems by pattern recognition—II, Bell System Technical Journal 40: 1–41 (1961).
\bibitem{berger} R. Berger, The undecidability of the domino problem, Memoirs of the American Mathematical Society 66: 1–72 (1966).
\bibitem{robinson} R. M. Robinson, Undecidability and nonperiodicity for tilings of the plane. Invent. Math. 12: 177-209 (1971).
\bibitem{penrose} R. Penrose, The role of aesthetics in pure and applied mathematical research, Bull. Inst. Math. Applications 10: 266-271 (1974).

\bibitem{hat} D. Smith, J. S. Myers, C. S. Kaplan, and C. Goodman-Strauss, An aperiodic monotile, Combinatorial Theory 4 (1) (2024).
\bibitem{spectre} D. Smith, J. S. Myers, C. S. Kaplan, and C. Goodman-Strauss, A chiral aperiodic monotile, Combinatorial Theory 4 (2) (2024).

\bibitem{Thue} A. Thue, Uber unendliche Zeichenreihen, Norske vid. Selsk. Skr. Mat. Nat. Kl. 7 1-22 (1906). Reprinted in “Selected mathematical papers of Axel Thue,” T. Nagell, ed., Universitetsforlaget, Oslo, pp. 139–158 (1997).
\bibitem{Morse} M. Morse, Recurrent geodesics on a surface of negative curvature, Trans. Amer. Math. Soc. 22: 84–100 (1921).
\bibitem{keane} M. Keane, Generalized Morse sequences, Z. Wahr. Verw. Geb. 10: 335-353 (1968).

\bibitem{sturmian1} M. Morse and G. A. Hedlund, Symbolic dynamics II. Sturmian trajectories. Am. J. Math. 62, 1–42 (1940).
\bibitem{sturmian2} J.-P Allouche and J. Shallit, Automatic Sequences: Theory, Applications, Generalizations. Cambridge University Press, 
Cambridge (2003).

\bibitem{mozes} S. Mozes, Tilings, substitution systems and dynamical systems generated by them, J. D'Analyse Mathematique 53: 139-186 (1989).

\bibitem{opterg} O. Jenkinson, Ergodic optimization in dynamical systems, Ergod. Theory Dyn. Syst. 39: 2593 - 2618 (2018).
\bibitem{typical} A. van Enter and J. Mi\c{e}kisz, Typical ground states for large sets of interactions, J. Stat. Phys. 181: 1906–1914 (2020). 
\bibitem{jaradin2} J. Mi\c{e}kisz, How low temperature causes long range order, J. Phys. A.: Math. Gen. 21: L529-L531 (1988).
\bibitem{jaradin3} J. Mi\c{e}kisz and C. Radin, 
Why solids are not really crystalline? Phys. Rev. B. 39: 1950-1952 (1989).

\bibitem{velenik} Sacha Friedli and Yvan Velenik, Statistical Mechanics of Lattice Systems - A Concrete Mathematical Introduction, Cambridge
University Press (2018).
\bibitem{quasireview} J. Mi\c{e}kisz, 
Quasicrystal problem -- on rigidity of non-periodic structures from statistical mechanics point of view,
Proceedings of the XLI Workshop on Geometric Methods in Physics, Bialystok, Poland, P. Kielanowski, A. Dobrogowska, D. Fernandez, and T. Goli\'{n}ski eds. Trends in Mathematics, Birkh$\ddot{a}$user (2025).

\bibitem{radinquasi} C. Radin, Crystals and Quasicrystals: A lattice gas model. Phys. Lett. 14A: 381--383 (1986).
\bibitem{manyphases} J. Mi\c{e}kisz, Many phases in systems without periodic ground states, Commun. Math. Phys. 107: 577-586 (1986).
\bibitem{sequenceoftemp} J. Mi\c{e}kisz, A microscopic model with  quasicrystalline properties, J. Stat. Phys. 58: 1137-1149 (1990).

\bibitem{simon} B. Simon, Fifteen problems in mathematical physics, Perspectives in mathematics: Anniversary of Oberwolfach 1984. Basel: Birkh$\ddot{a}$user (1984).

\bibitem{ban} M. Bundaru, N. Angelescu, and G. Nenciu, On the ground state of Ising chains with finite range interactions, Phys. Lett. 43A: 5--6 (1973).
\bibitem{radinschulman} C. Radin and L. Schulman, Periodicity of  classical ground states, Phys. Rev. Lett. 51: 621--622 (1983).
\bibitem{thirdlaw} J. Mi\c{e}kisz and C. Radin, 
The Third Law of  thermodynamics, Mod. Phys. Lett. B1: 61-65 (1987).

\bibitem{sbp1} J. Mi\c{e}kisz, Stable quasicrystalline ground states,
J. Stat. Phys. 88: 691--711 (1997).
\bibitem{sbp2} J. Mi\c{e}kisz, Classical lattice-gas models of quasicrystals, J. Stat. Phys. 97: 835--850 (1999).

\bibitem{peyriere} J. Peyriere, Frequency of patterns in certain graphs and in Penrose tilings, J. Phys. Colloq. 47: C 41--62 (1986).
\bibitem{gambaudo} J. Aliste-Prieto, D. Coronel, and J.-M. Gambaudo, Rapid convergence to frequency for substitution tilings of the plane, 
Commun. Math. Phys. 306: 365--380 (2011).
\bibitem{aizenman} M. Aizenman, S. Goldstein, 
and J. L. Lebowitz, Bounded fluctuations and translation symmetry breaking in one-dimensional particle systems, J. Stat. Phys. 103: 601-618 (2000).

\bibitem{torquato1} E. C. Oguz, J. E. S. Socolar, P. J. Steinhardt, and S. Torquato, Hyperuniformity of quasicrystals, Phys. Rev. B 95:  054119 (2017).
\bibitem{torquato2} S. Torquato, Hyperuniform states of matter, Physics Reports 745: 1-95 (2018).
\bibitem{maher} C. Maher and S. Torquato, Local order metrics for many-particle systems across length scales, Phys. Rev. Research 6: 033262 (2024).

\bibitem{unstable} J. Mi\c{e}kisz and C. Radin, The unstable chemical structure of quasicrystalline alloys, Phys. Lett. 119A: 133-134  (1986).
\bibitem{gonsquas} J. S. Gonschorowski, A. Quas, and J. Stefken, Support stability of maximizing measures for shifts of finite type, Ergod. Theory Dyn. Syst. 41: 869–880 (2021).

\bibitem{ghmorse} W. H. Gottschalk and G. A. Hedlund, A characterization of the Morse minimal set, Proc. Am. Math. Soc. 15: 70-74 (1964).
\bibitem{tmhamiltonian} C. Gardner, J. Mi\c{e}kisz, C. Radin, and A. van Enter, Fractal symmetry in an Ising model, J. Phys. A.:  Math. Gen. 22: L1019--L1023 (1989).

\bibitem{sturmiangs} A. van Enter, H. Koivusalo, and J. Mi\c{e}kisz, Sturmian ground states in classical lattice–gas models, 
J. Stat. Phys. 178:  832–844 (2020).
\bibitem{jspnonstab} D. Glodkowski and J. Mi\c{e}kisz, On non-stability of one-dimensional non-periodic ground states, J. Stat. Phys. 192: 4 (2025).

\bibitem{sadun} F. Adiceam, D. Damanik, F. Gähler, U. Grimm, A. Haynes, A. Julien, A. Navas, L. Sadun, and B. Weiss, Open problems and conjectures related to the theory of mathematical quasicrystals, 
Arnold Math. J. 2: 579–592 (2016).
\bibitem{baake} M. Baake and U. Grimm, Aperiodic Order, Volume 1:
A  Mathematical Invitation, Cambridge University Press (2013).

\end{thebibliography}
\end{document}